\documentstyle[seceq,twocolumn,epsf]{jpsj}

\def\H{{\cal H}}
\def\HSS{{\rm HSS}}
\def\UTS{{\rm UTS}}
\def\DTS{{\rm DTS}}
\def\u{{\rm u}}
\def\d{{\rm d}}
\def\h{{\rm h}}
\def\epg{\epsilon_{\scriptsize\rm g}}
\def\mpynn{$m$-MPYNN$\cdot$BF$_4$}
\def\Hu{{\cal H}_{\scriptsize \rm u}}
\def\Hd{{\cal H}_{\scriptsize \rm d}}

\def\simgeq{\mbox{\raisebox{-1.0ex}{$\stackrel{>}{\sim}$}}}

\def\v#1{\mib #1}

\def\runtitle{Ground State and Excitations of the $S=1$ Kagom\'e Antiferromagnet
}

\def\runauthor{Kazuo {\sc Hida}}

\title
{
Ground State and  Elementary Excitations of the $S=1$ Kagom\'e Heisenberg Antiferromagnet
}

\author
{Kazuo {\sc Hida}
\footnote{e-mail: hida@phy.saitama-u.ac.jp}}

\inst
{
Department of Physics, Faculty of Science,\\ Saitama University, Urawa, Saitama 338-8750
}

\recdate
{
\today
}

\abst
{
Low energy spectrum of the $S=1$ kagom\'e Heisenberg antiferromagnet (KHAF) is studied by means of exact diagonalization and the cluster expansion. The magnitude of the energy gap of the magnetic excitation is consistent with the recent experimental observation for \mpynn. In contrast to the $S=1/2$ KHAF, the non-magnetic excitations have finite energy gap comparable to the magnetic excitation. As a physical picture of the ground state, the hexagon singlet solid state is proposed and verified by variational analysis.}

\kword
{
kagom\'e Heisenberg antiferromagnet, Pad\'e approximation, cluster expansion method, numerical diagonalization, hexagonal singlet solid state
}

\begin{document}
\maketitle


\section{Introduction}

The quantum antiferromagnets on the kagom\'e lattice has been extensively studied theoretically and experimentally because of the interest in the interplay of the strong quantum fluctuation and the highly frustrated nature of the lattice structure. So far, most of the attempts has been focused on the $S=1/2$ case.\cite{ze1,ey1,nm1,wal1} Owing to these extensive studies, it is widely accepted that the ground state of this model is a spin singlet state and the triplet excitation has a finite energy gap. Furthermore, there are a number of singlet excitataions below the first triplet excitation possiblly down to zero energy in the thermodynamic limit.\cite{wal1}
\begin{figure}
\epsfxsize=70mm 
\centerline{\epsfbox{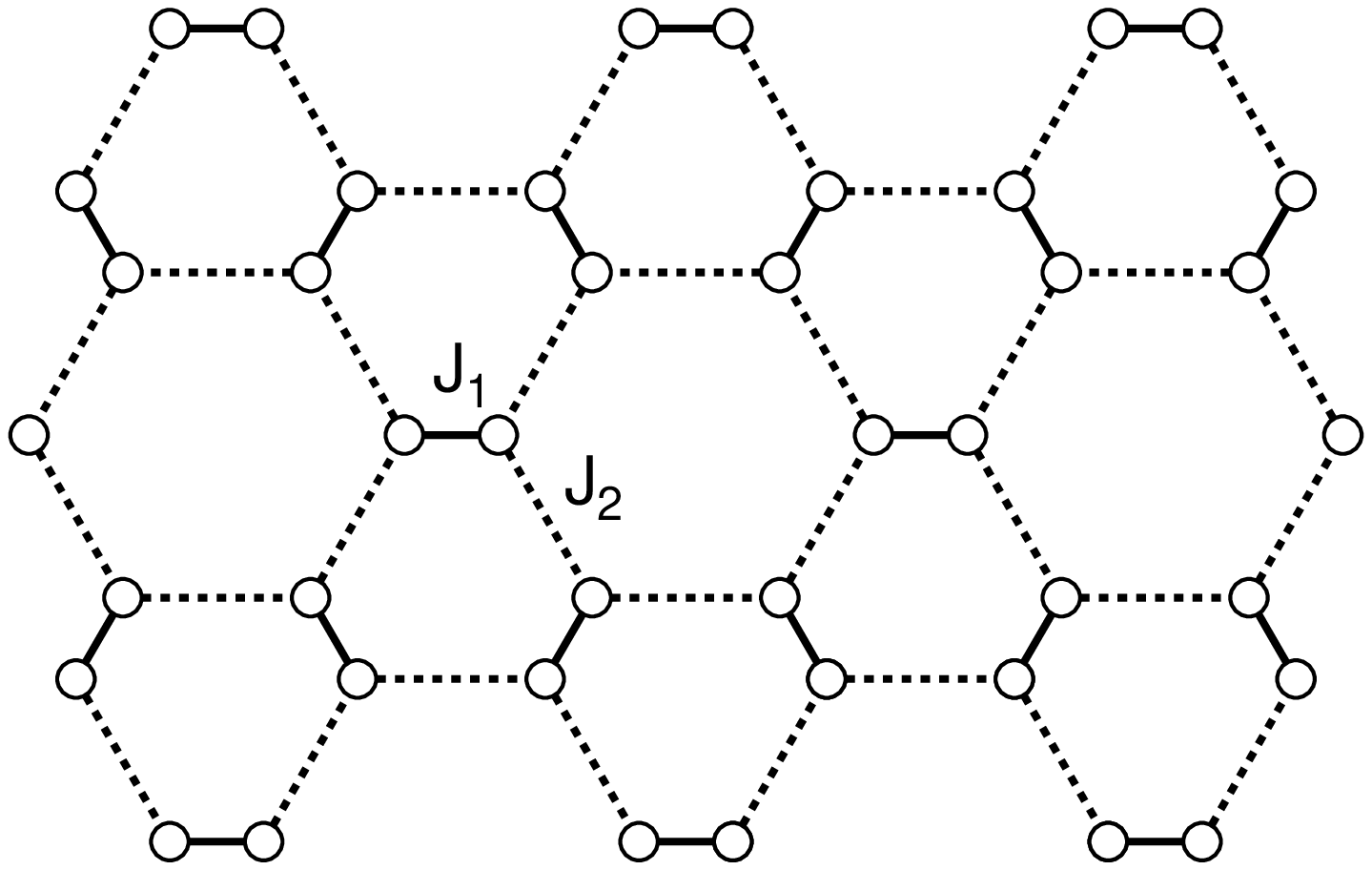}}
\caption{The structure of \mpynn lattice.}
\label{fig1}
\end{figure}

Recently, Wada and coworkers\cite{wada1,agawa1,wata1} have synthesized the material  \mpynn  and measured its magnetic and thermal properties. This material has the two dimensional structure which is schematically depicted in Fig. \ref{fig1}. Magnetically it can be regarded as the Heisenberg model consisting of dimers of two $S=1/2$ spins on each {\it bond} of the triangular lattice. The intradimer exchange coupling $J_1$ is ferromagnetic and depicted by the thick solid lines. These spins are also coupled via  interdimer antiferromagnetic couplings $J_2$ represented by the thick dotted lines\cite{agawa1,wada1,wata1}. Thus the Hamiltonian is given by,
\begin{equation}
\label{ham0}
H = J_1 \sum_{\scriptsize \rm intradimer} \v{s}_{i} \v{s}_{j} + J_2 \sum_{\scriptsize \rm interdimer} \v{s}_{i} \v{s}_{j} 
\end{equation}
where  $\v{s}_{i}$ is the spin-1/2 operartor. Experimentally, the exchange constants are estimated as $J_1 \simeq -23.26$K and $J_2 \simeq 3.11$K from susceptibility measurements, so that  $\mid J_1 \mid >> \mid J_2 \mid $. Therefore, at low temperatures, this material can be regarded as the spin-1 kagom\'e Heisenberg antiferromagnet (KHAF) described by the following Hamiltonian,

\begin{equation}
\label{ham1}
H = J \sum_{<i,j>} \v{S}_{i} \v{S}_{j}
\end{equation} 
where  $\v{S}_{i}$ is the spin-1 operator and the summation is taken over all nearest neighbour pairs of sites of the kagom\'e  lattice. The effective exchange constant between $S=1$ spins is given by $J = J_2/4 \simeq 0.78$K. Wada et al.\cite{wada1} confirmed that the magnetic excitation of this material has a finite energy gap $\Delta \simeq 0.25$K from low temperature susceptibility measurement.

Theoretically, Asakawa and Suzuki\cite{as1} have carried out the cluster expansion calculation for the ground state energy up to the fourth order in $\lambda$ in their pioneering work. They calculated, however, the excitation gap only up to the first order so that conclusive result for the energy gap was not obtained.

In the present work, we investigate the ground state and excitation spectrum of this model (\ref{ham1}) by the numerical diagonalization and series expansion method. We also propose a possible physical picture of the ground state named hexagon singlet solid (HSS) state. To confirm this picture, we perform the variational analysis of the ground state energy based on the HSS wave function.

This paper is organized as follows: In \S 2, the model Hamiltonian is presented. The numerical method employed in this work is explained in \S 3. The numerical results are presented in \S 4. The results are compared with experimental value for \mpynn . The HSS picture of the ground state is presented in \S 5. The last section is devoted to a summary and discussion.

\section{Model Hamiltoninan}

\begin{figure}
\epsfxsize=60mm 
\centerline{\epsfbox{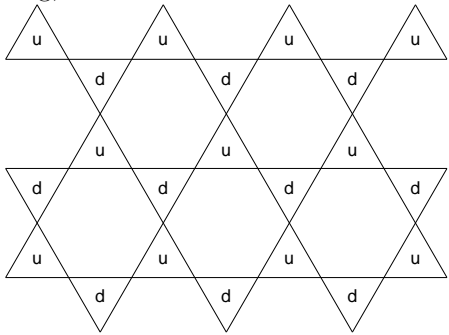}}
\caption{Structure of the kagom\'e lattice. The UTC's and DTC's are denoted by u and d, respectively.}
\label{fig2}
\end{figure}

Let us rewrite the Hamiltonian of $S=1$ KHAF as follows,

\begin{eqnarray}
\label{ham2}
\H &=& \Hu + \lambda\Hd, \\
\Hu &=& J \sum_{\bigtriangleup}\v{S}_{i} \v{S}_{j}, \\
\Hd &=& J \sum_{\bigtriangledown}\v{S}_{i} \v{S}_{j}, 
\end{eqnarray} 
where $\sum_{\bigtriangleup}$ and  $\sum_{\bigtriangledown}$ represent the summation over the bonds in the upward pointed triangular clusters (UTC) and downward pointed triangular clustrs (DTC), respectively, which are depicted in Fig. \ref{fig2}.  Although we are interested in the uniform KHAF, we introduce the trimerization parameter $\lambda$ for the computational purpose. In the following, we take the unit $J=1$.

\section{Numerical Methods}
\subsection{Exact Diagonalization}

\begin{figure}
\epsfxsize=70mm 
\centerline{\epsfbox{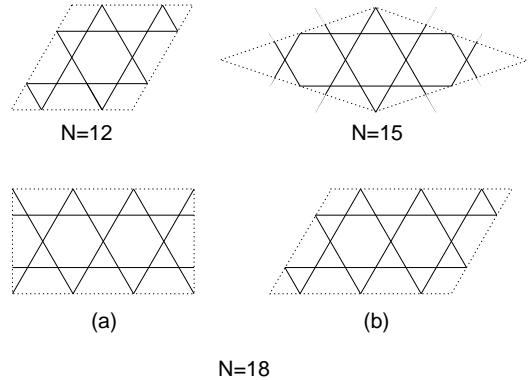}}
\caption{The clusters used for numerical diagonalization.}
\label{cluster}
\end{figure}

For $N=12$, 15 and $18$, the excitation energies of the low lying singlet and triplet excited states are calculated by the numerical diagonalization method using the Lanzcos algorithm. The clusters shown in Fig. \ref{cluster} are used with periodic boundary condition. 

\subsection{Cluster Expansion Method}

As a complementary method, the magnetic excitation with $S=1$ is calculated using the cluster expansion method treating  $\Hd$ as a perturbation. 

Following Asakawa and Suzuki\cite{as1}, we take $\lambda=0$ in the unperturbed state. In this case, the ground state is an simple assembly of independent singlet states on UTC. In the following, this state is called upward pointed triangular singlet (UTS) state and the corresponding state on  DTC's is called downward pointed triangular singlet (DTS) state .  We apply the series expansion with respect to $\lambda$ using the method of Gelfand, Singh and Huse\cite{sgh1,gsh1} and Gelfand\cite{gel1} to obtain the series for the triplet excitation energy $\epsilon(\v{k})$ up to the third order in $\lambda$ where $\v{k}$ is the wave vector associated with the excitation. 
\begin{equation}
\epsilon(\v{k};\lambda) = \sum_{m=0}^{M} c_m (\v{k}){\lambda}^m \ \ \mbox{with} \ M=3. 
\label{raw}
\end{equation}
The energy gap evaluated from this series is in itself ill-behaved for $\lambda \simgeq 0.3$. We therefore employ the Pad\'e approximant,
\begin{equation}
\epsilon(\v{k};\lambda)^{[K,L]} = \frac{\sum_{k=0}^{K}p_k (\v{k}) {\lambda}^k}{\sum_{l=0}^{L}q_l (\v{k}) {\lambda}^l}, 
\label{pade}
\end{equation}
with $p_0 (\v{k}) = 1$. 

In the usual Pad\'e approximation, $K+L$ must be equal to $M$. The original Hamiltonian (\ref{ham2}), however, has the symmetry $H(1/\lambda) = H(\lambda)/\lambda$. Therefore the energy gap has also the same symmetry $\epsilon(\v{k};1/\lambda)=\epsilon(\v{k};\lambda)/\lambda$. We therefore impose the same symmetry for the Pad\'e approximant (\ref{pade}) which ensures $p_k=p_{K-k}$  and $q_l=q_{L-l}$ with $K=L+1=M$. In the following, we call this approximation as 'symmetric Pad\'e approximation'. We also applied this approximation to the 4-th order series of the ground state energy per site $\epg$ obtained by Asakawa and Suzuki\cite{as1} which yields the value $\epg=-1.4468$. This is close to the values estimated from exact diagonalzation of $N=18$ clusters, namely $\epg=-1.4511$ (cluster(a)) or $-1.4393$ (cluster(b)). This confirms the reliability of symmetric Pad\'e approximation.  

\section{Numerical Results}

\begin{figure}
\epsfxsize=70mm 
\centerline{\epsfbox{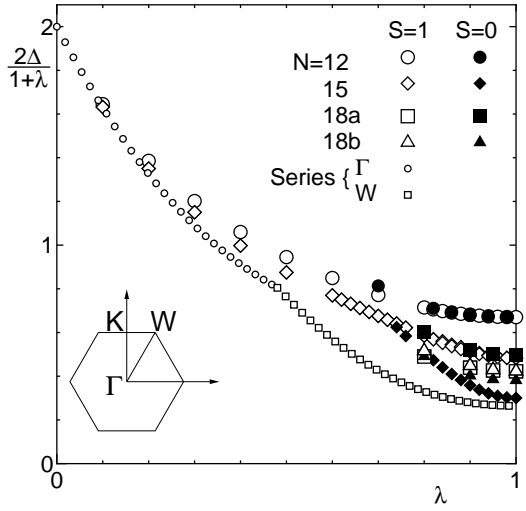}}
\caption{The $\lambda$ dependence of the energy gaps. The small symbols are the estimation from the symmetric Pad\'e approximant of the cluster expansion series. The large symbols are the exact diagonalization results for $N=12$, 15 and 18, respectively. The inset is the Brillouin zone of the kagom\'e lattice. Energy gaps are normalized by a factor $(1+\lambda)/2$ so that the symmetry $\Delta(1/\lambda) = \Delta(\lambda)/\lambda$ is evident.}
\label{fig4}
\end{figure}

\begin{figure}
\epsfxsize=70mm 
\centerline{\epsfbox{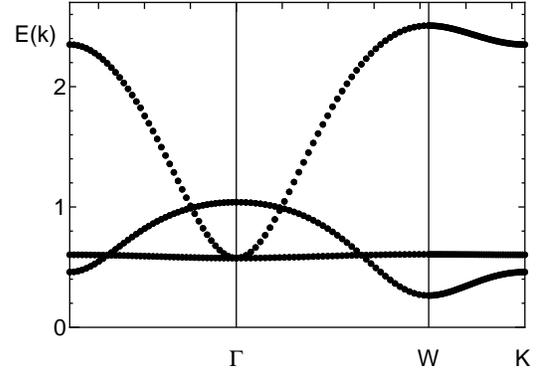}}
\caption{The dispersion relation of $S=1$ KHAF estimated from the symmetric Pad\'e approximant of the cluster expansion series. }
\label{disp}
\end{figure}

The $\lambda$-dependence of the energy gaps is shown in Fig. \ref{fig4}. According to the diagonalization results, the ground state is always the singlet state with total spin $S=0$. The $\lambda$-dependence of the energy gap is common to all diagonalization data and series expansion data for small $\lambda$. For small $\lambda$, the lowest excitation has always $S=1$. This corresponds to the triplet excitation localized in UTC. As $\lambda$ increases, however, the properties of the lowest excited states vary depending on the methods used, size and form of the clusters. 
  According to the diagonalization data, the singlet excitation with $S=0$ comes down as $\lambda$ approaches unity. At $\lambda=1$, the lowest excitation has $S=1$ or $S=0$ depending on the system size and forms of the cluster. They are summarized in table \ref{table1}. The first excited state has $S=1$ for $N=12$ and type (a) cluster with $N=18$. On the other hand, for $N=15$ and type (b) cluster with $N=18$, the first excited state is $S=0$ non-magnetic state. The magnetic gap and non-magnetic gap are close to each other for $N=18$. As for the wave number of the excitations, most of the exact diagonalization data shows that the lowest excited state has wave number $\v{k}=0$ ($\Gamma$-point) except for the case of type (a) $N=18$ cluster for which the magnetic gap is located at the K-point near $\lambda=1$. On the other hand,  the series expansion result shows that the magnetic gap is located at the W-point of the Brillouin zone for $\lambda > 0.48$ while it is located at the origin for $\lambda < 0.48$. The dispersion relation estimated from the series expansion is shown in Fig. \ref{disp}.


\begin{figure}
\epsfxsize=70mm 
\centerline{\epsfbox{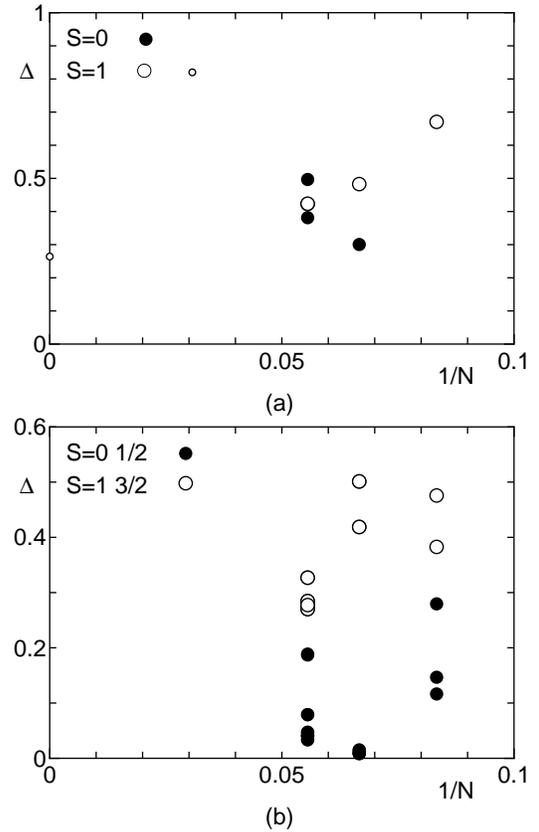}}
\caption{The system size dependence of the energy gaps for $\lambda=1$ for (a) $S=1$ and (b) $S=1/2$ KHAF. The small symbol in (a) is the estimation from the symmetric Pad\'e approximant of the cluster expansion series. The large symbols are the exact diagonalization results. In (a), for $N=12$ the filled circle is almost covered by the open circle.}
\label{fig6}
\end{figure}
\vspace{5mm}

\begin{table}
\caption{Energy gap, spin quantum numbers and wave numbers of the low lying excited states of finite size clusters. The values with * are the lowest excitations for each cluster.}
\label{table1}
\begin{tabular}{@{\hspace{\tabcolsep}\extracolsep{\fill}}clclc} \hline
$N$ & $\Delta (S=0)$ &  $k$ & $\Delta (S=1)$ &  $k$\\
\hline
\hline
12 & 0.67074 & 0 & 0.67051*  & 0 \\
\hline
15 &0.30050* & 0 & 0.48270  & 0 \\
\hline
18a & 0.49656  & 0  & 0.42262* & K-point\\
\hline
18b &  0.38134* & 0 & 0.42343 & 0 \\
\hline
\end{tabular}
\end{table}

To make clear the difference between the excitation spectra of $S=1/2$ and $S=1$ KHAF, the system size dependence of the gaps are shown in Fig. \ref{fig6}(a) and (b) for $S=1$ and $S=1/2$ cases, respectively. For $S=1/2$ KHAF, the non-magnetic gap is much smaller than the magnetic gap. Actually, it is expected that the low energy non-magnetic excitations are distributed quasi-continuously down to zero according to the detailed calculation with larger clusters\cite{nm1,wal1}. For $S=1$ case, however, both gaps have the same order of magnitude within the system sizes studied here and we find no indication of the appearance of the quasi-continous non-magnetic excitation spectrum.

In any case, the magnitude of the magnetic gap is estimated between $0.42J$ ($N=18$ cluster) and $0.26J$ (series expansion). The magnetic gap for \mpynn is estimated as 0.25K from low temperature susceptibility measurement. This value corresponds to $0.32J$ which is consistent with present estimation.

\section{Hexagon Singlet Solid Picture}

The presence of finite magnetic gap suggests that the ground state consists of an assembly of localized singlet states. Apparently, there are three candidates of such singlet states. Two of them are UTS and DTS states which are the ground states for $\lambda=0$ or $\lambda \rightarrow \infty$, respectively. At $\lambda=1$, however, the symmetry betwenn UTC's and DTC's is recovered. The ground state therefore should contain both states with equal weight. 

\begin{figure}
\epsfxsize=70mm 
\centerline{\epsfbox{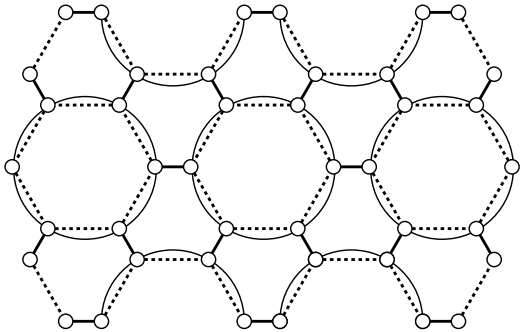}}
\caption{The hexagon singlet solid state. The six $S=1/2$ spins on a circle form a hexgaon singlet state. The pairs of $S=1/2$ spins connected by the thick solid lines are symmetrized to yield $S=1$ spins. }
\label{fig7}
\end{figure}

There is another possibility that the singlet states are formed on the hexgaonal clusters. Especially, if we decompose a $S=1$ spin into two $S=1/2$ spins, as in the VBS model of Haldane gap state\cite{aklt}, we can construct a trial state covered by local singlets without breaking the UTC-DTC symmetry as depicted in Fig. \ref{fig7}. Hereafter we call this state as hexagon singlet solid (HSS) state.

The wave function of the HSS state can be written down by expressing the single $S=1$ spin state as,
\begin{eqnarray}
&&\mid 1 >=\psi_{1/2,1/2}/\sqrt{2} \\ \nonumber
&&\mid 0 >=\psi_{1/2,-1/2}\equiv \psi_{-1/2,1/2} \\ \nonumber
&&\mid -1 >=\psi_{-1/2,-1/2}/\sqrt{2}
\end{eqnarray}
where $\psi_{\alpha,\beta}$ is the basis state of the $S=1$ spin in terms of the two symmetrized $S=1/2$ states defined as,
\begin{equation}
\psi_{\alpha,\beta}=[\psi_{\alpha}\otimes\psi_{\beta}+\psi_{\beta}\otimes\psi_{\alpha}]/\sqrt{2}.
\end{equation}
Here $\psi_{\alpha}$ is the basis state of the $S=1/2$ spin.

Following the construction of the VBS state by Affleck et al\cite{aklt}, we construct the HSS state as,

\begin{eqnarray}
\Phi^{\scriptsize \HSS}&=& \bigotimes_i \psi_{\alpha_i,\beta_i}\prod_i (\delta_{\alpha_i,\gamma_i}+\delta_{\beta_i,\gamma_i}) \nonumber \\ 
&\times&\prod_p w^{\gamma_{i_p}\gamma_{j_p}\gamma_{k_p}\gamma_{l_p}\gamma_{m_p}\gamma_{n_p}}
\end{eqnarray}
where the summation is assumed for the repeated indices. The $i$-th site has two $S=1/2$ indices $\alpha_{i}$ and $\beta_{i}$. The index $p$ distingush the hexagons and the sites $\{i_p,j_p,k_p,l_p,m_p,n_p\}$ belong to the $p$-th hexagon. The wave function $w^{\gamma_{i_p}\gamma_{j_p}\gamma_{k_p}\gamma_{l_p}\gamma_{m_p}\gamma_{n_p}}$ is that of the lowest singlet state of a single  hexagon composed of six $S=1/2$ spins. 

Taking into account the two triangular singlet states $\Phi^{\scriptsize \UTS}$ and $\Phi^{\scriptsize \DTS}$, we take the variational state as

\begin{equation}
\label{var}
\Phi = c_{\u} \Phi^{\scriptsize \UTS} + c_{\d} \Phi^{\scriptsize \DTS} + c_{\h}\Phi^{\scriptsize \HSS}
\end{equation}
and minimize the energy expectation values for two types of clusters with $N=18$ for $\lambda=1$. As shown in table \ref{table2}, $c_{\scriptsize \u}, c_{\scriptsize \d}  << c_{\scriptsize \h}$ and the variational ground state energy is close to the exact value for both clusters. If we only admit two triangular singlet state, the variational energy per site is almost equal to $-1$ which is far from the exact value. On the other hand, if we take only the HSS state the variational energy is around $-1.3$ for both clusters. This value is already very close to the optimum value. Thus the HHS picture gives a reasonable physical picture of the ground state of $S=1$ KHAF. 
\begin{fulltable}
\caption{Various estimations of ground state energy per site $\epg$ for $\lambda=1$. The values estimated from variational function (\ref{var}), pure HSS state, UTS+DTS state, numerical diagonalization and series expansion are tarbulated. For the variational estimation, optimum values for the coefficients of HSS, UTS and DTS states are also shown.}
\label{table2}
\begin{fulltabular}{@{\hspace{\tabcolsep}\extracolsep{\fill}}cccccccc} \hline
Cluster & $\epg$(\HSS+\UTS+\DTS)&  $c_{\h}$ &  $c_{\u} (= c_{\d})$ & $\epg$(\HSS) & $\epg$(\UTS+\DTS) & $\epg$(exact) &  $\epg$(series) \\
\hline
\hline
18(a) & $-1.3031$ & 0.9921 & 0.0326  &$-1.3028$ &$-1.0015$ &  $-1.4511$ & {\raisebox{-2.0ex}[0pt]{$-1.4468$}} \\
\cline{1-7}
18(b) & $-1.3081$ & 0.9966 & 0.0168  & $-1.3080$ &$-1.0010$ &  $-1.4393$ & \\
\hline
\end{fulltabular}
\end{fulltable}

\section{Summary and Discussion}

The spin-1 kagom\'e lattice antiferromagnetic Heisenberg model is studied by means of the cluster expansion and exact diagonalization method. The quantum numbers of the lowest excited state depends crucially on the method used, size and form of the clusters. This, in turn, confirms that the excitation spectrum of KHAF is strongly degenerate even for $S=1$ case. It is suggested that the nonmagnetic excitations have finite energy gap as well as the magnetic excitations. The HSS picture of the ground state is presented and varified by the variational argument. 

The material \mpynn, which motivated the present study, can be regarded as the assembly of $S=1/2$ spins on the hexagon as shown in Fig. \ref{fig1}.  If the interhexagon coupling $J_1$ vanishes, the ground state is the six spin singlet state on each hexagon. If $J_1$ is switched on, the complete singlet state is perturbed. As far as  $J_1$ is ferromagnetic, however, it cannot kill the spin degrees of freedom totally. Actually, we know that the $S=1/2$ antiferromagnetic-ferromagnetic alternating Heisenberg chain has an energy gap even in the limit of infinitely strong ferromagnetic bond giving rise to the Haldane phase of the $S=1$ Heisenberg chain.\cite{kh1} From this analogy, also, it is reasonable that the magnetic gap remains finite in the present material even in the limit of $J_1 \rightarrow -\infty$ which is nothing but the $S=1$ KHAF.

 On the basis of HSS picture, we can also understand the physical origin of the finite non-magnetic gap in the following way. There are local singlet excited states in each single $S=1/2$ hexagon. These states have the exciation energy of the same order of magnitude as the magnetic excitaions. If $J_1$ is switched on, these states will tend to a gapped non-magnetic excitations.

The author thanks N. Wada for fruitful discussion and comments. The numerical calculation is performed using the FACOM VPP500 and HITAC SR8000 at the Supercomputer Center, Institute for Solid State Physics, the University of Tokyo,  HITAC SR8000 at the Information Technology Center, the University of Tokyo and the HITAC S820/80 at the Information Processing Center, Saitama University.  This work is supported by the Grant-in-Aid for Scientific Research from the Ministry of Education, Science, Sports and Culture.

\end{document}